\documentclass[namedreferences]{solarphysics}
\usepackage[optionalrh,solaenum]{spr-sola-addons} 
\usepackage{graphicx}                    
\usepackage{color}                       
\usepackage{url}                         
\usepackage{rotating}

\begin{document}

\begin{article}

\begin{opening}

\title{How to optimize nonlinear force-free coronal magnetic
field extrapolations from SDO/HMI vector magnetograms?}

%
\author{T.~\surname{Wiegelmann} $^{1}$\sep
J.K. ~\surname{Thalmann} $^{1}$\sep
B. ~\surname{Inhester} $^{1}$\sep
T. ~\surname{Tadesse} $^{1,3}$\sep
X. ~\surname{Sun} $^{2}$\sep
J.T. ~\surname{Hoeksema} $^{2}$ 
       }

%
\runningauthor{Wiegelmann et al.}
\runningtitle{NLFFF with SDO/HMI}

%
  \institute{$^{1}$ Max-Planck-Institut f\"ur Sonnensystemforschung,
Max-Planck-Strasse 2, 37191 Katlenburg-Lindau, Germany
                     email: \url{wiegelmann@mps.mpg.de}\\ 
            $^{2}$ W.W. Hansen Experimental Physics Laboratory,
            Stanford University, Stanford, CA 94305, USA \\
            $^{3}$ Addis Ababa University, College of Natural Sciences,
            Institute of Geophysics, Space Science, and Astronomy,
            Po.Box 1176,  Addis Ababa, Ethiopia \\
             }

\begin{abstract}
$ $ \\
The SDO/HMI instruments provide  photospheric vector
magnetograms with a high spatial and temporal
resolution. Our intention is to model
the coronal magnetic field above active regions with the help of a
nonlinear force-free extrapolation code. Our code is based on an optimization
principle and has been tested extensively with semi-analytic and numeric
equilibria and been applied before to  vector magnetograms from
Hinode and ground based observations. Recently we implemented a new version
which takes measurement errors in photospheric vector magnetograms into
account. Photospheric field measurements are often due to measurement errors
and finite nonmagnetic forces inconsistent as a boundary for
a force-free field in the corona. In order to deal with these uncertainties,
we developed two improvements: 1.) Preprocessing of the surface measurements
in order to make them compatible with a force-free field 2.) The new code keeps a
balance between the force-free constraint and deviation from the photospheric
field measurements.
Both methods contain free parameters, which have to be
optimized for use with data from SDO/HMI. Within this work we describe the
corresponding analysis method and evaluate the force-free equilibria by
means of how well force-freeness and solenoidal conditions are fulfilled,
the angle between magnetic field and electric current and by comparing
projections of magnetic field lines with coronal images from SDO/AIA.
We also compute the available free magnetic energy and discuss the potential
influence of control parameters.
\end{abstract}
%
%
\end{opening}
\section{Introduction}
The Helioseismic and Magnetic Imager (HMI) on board of the Solar
Dynamics Observatory (SDO) provides us with measurements based on
which the photospheric magnetic field vector can be derived
 \cite{schou:etal11}. Within this work we describe how these
measurements can be extrapolated into the solar corona under the
assumption that the coronal magnetic field is force-free, which
means that the Lorentz-force vanishes. We compare the resulting
magnetic field models with observations of the coronal plasma from
the Atmospheric Imaging Assembly (AIA), which is also onboard of
SDO.

\begin{figure}
\mbox{\includegraphics[width=6cm,clip=]{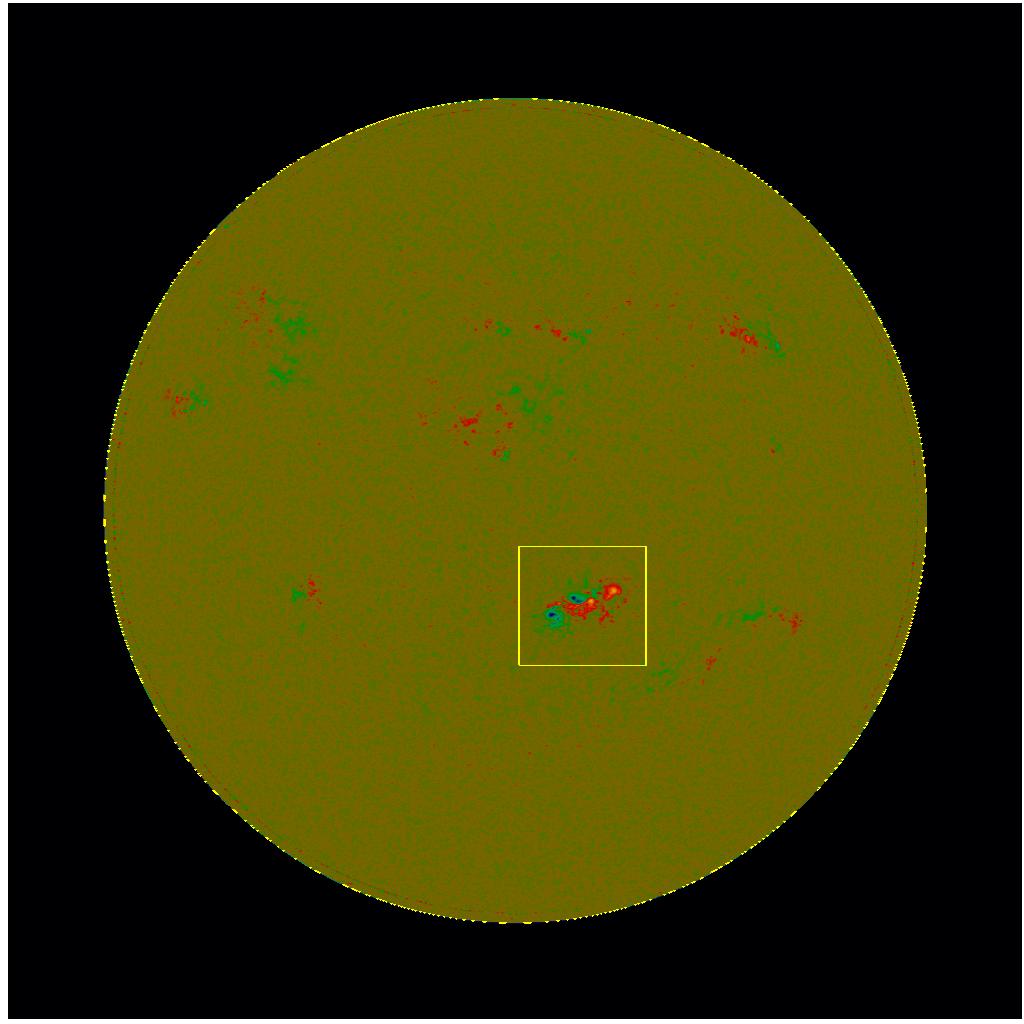}
\includegraphics[width=6cm,clip=]{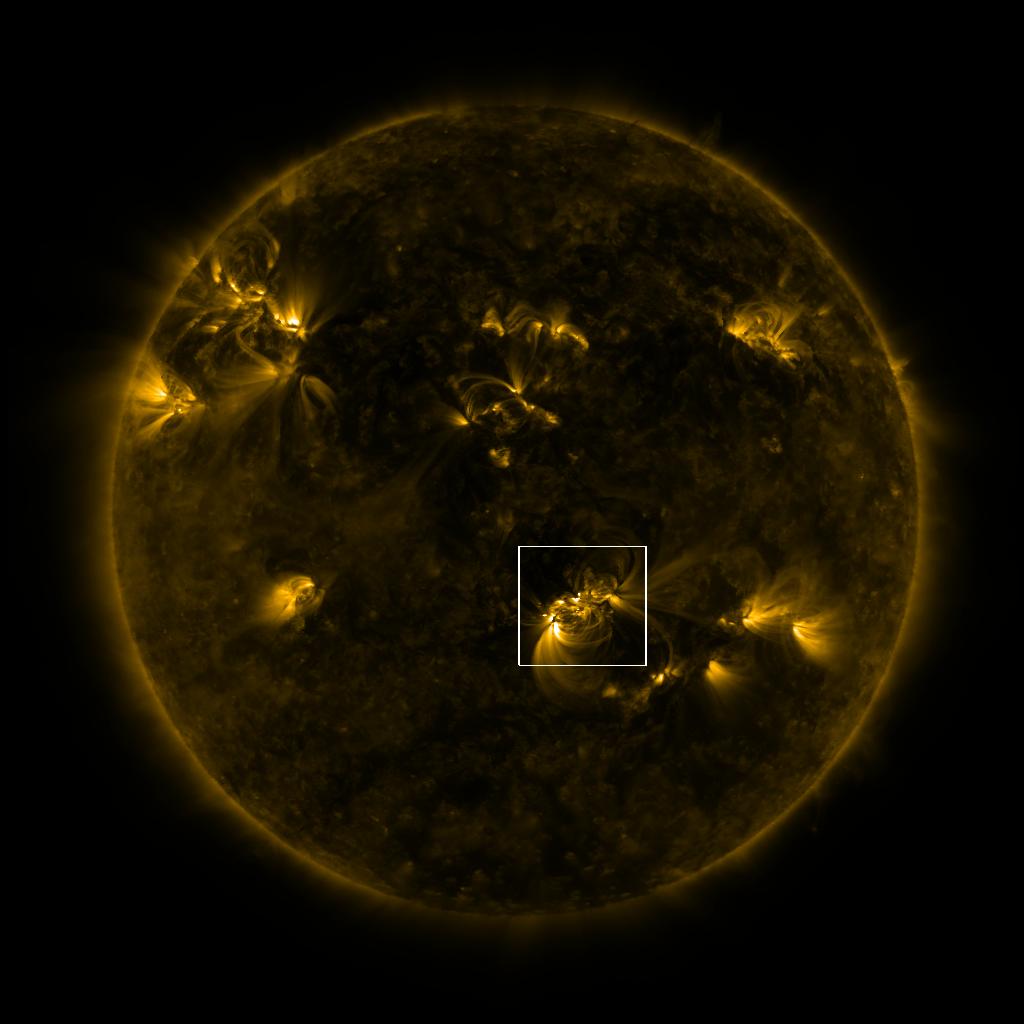}}
\caption{Left: Full disk SDO/HMI magnetogram, Right: Full disk
${\rm AIA} 171 \AA \,$ image. Both data sets have been obtained
on Feb. 14. 2011 at 20:34 and have been aligned
as described in the SDO data analysis guide
[DeRosa \& Slater 2011,
\url{http://www.lmsal.com/sdouserguide.html}].
The rectangle outlines the sub-region (AR11158) in the vector magnetogram
which is used for the force-free field modeling.
} \label{figure1}
\end{figure}

\begin{figure}
\mbox{\includegraphics[width=6cm,clip=]{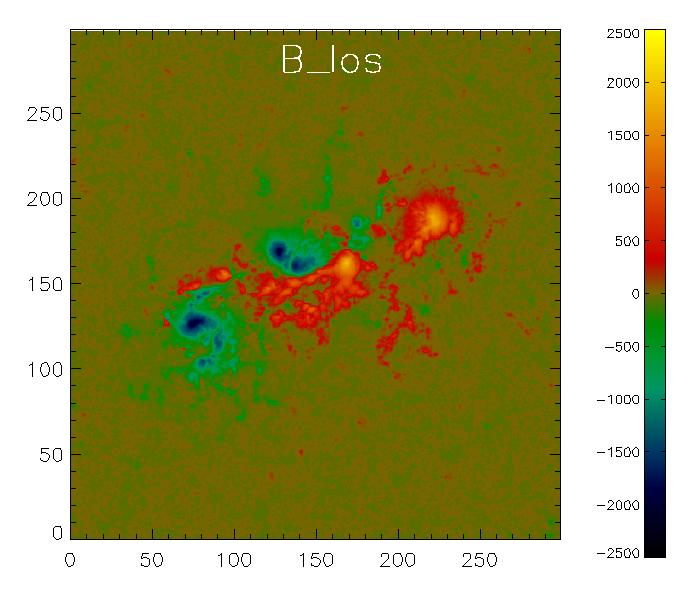}
\includegraphics[width=6cm,clip=]{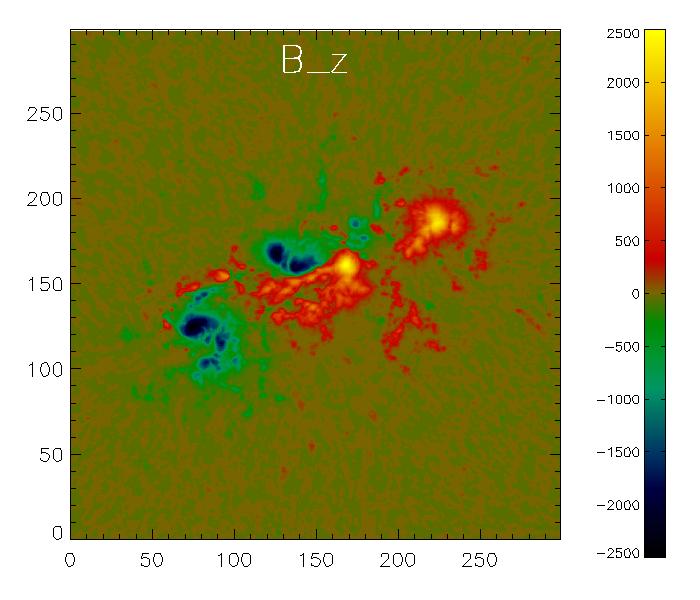}}
\mbox{\includegraphics[width=6cm,clip=]{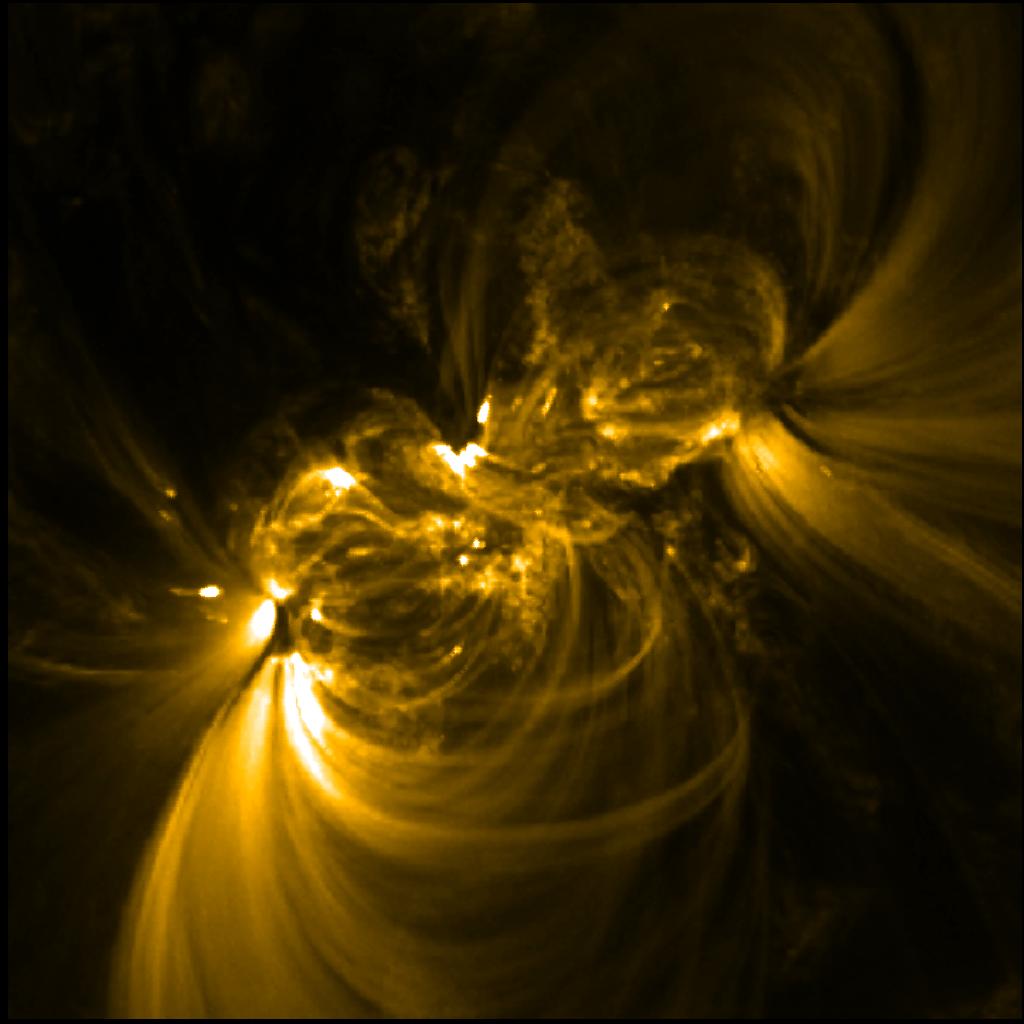}
\includegraphics[width=6cm,clip=]{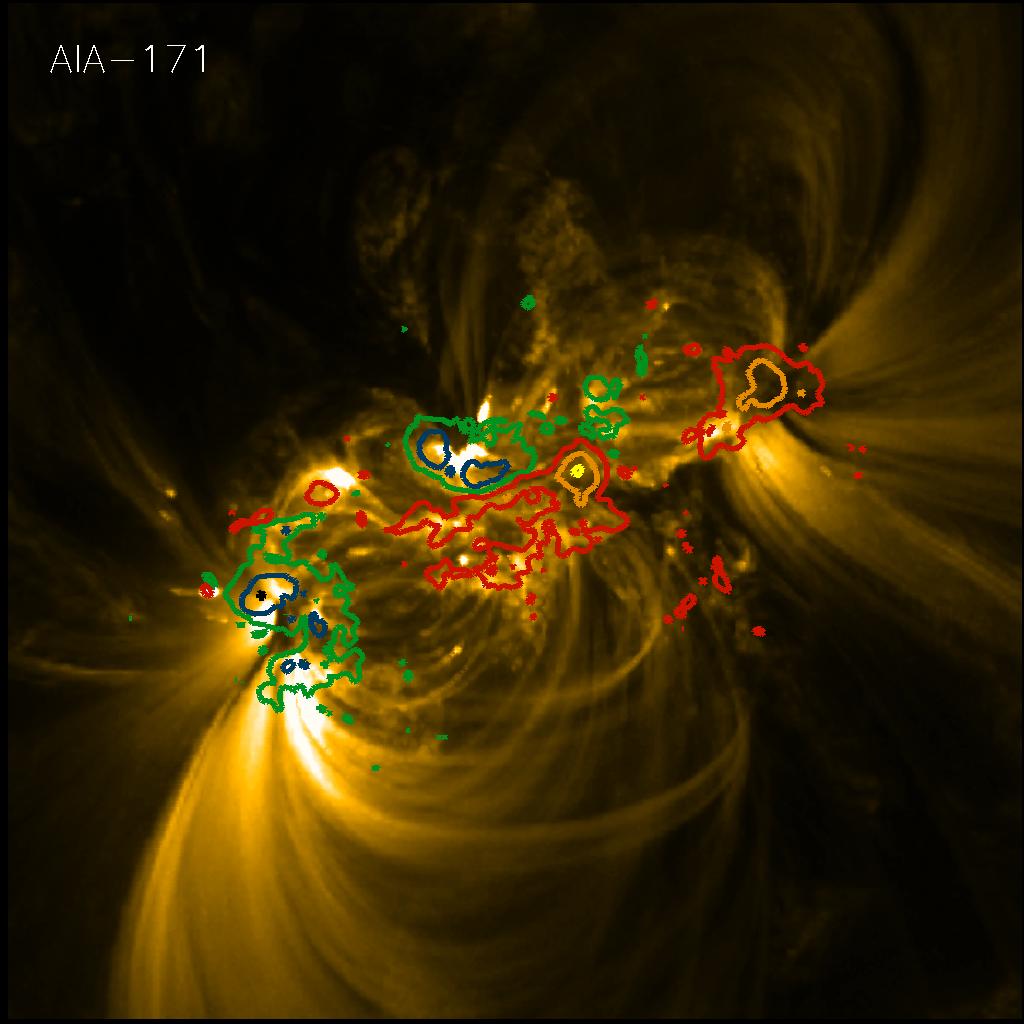}}
\caption{Top left: $B_{\rm los}$ cut from full disk HMI magnetogram.
Top right: $B_z$ from vector magnetogram. To aline the vector magnetogram
with the line of sight magnetogram and AIA we carried out a correlation
analysis and the correlation between $B_{\rm los}$ and $B_z$ is
$92 \%$.
Bottom: Same field of view seen in ${\rm AIA} 171 \AA$. In the right
picture we over-plotted contour lines of $B_z$
(same color-code as in images above) }
\label{figure2}
\end{figure}

\begin{figure}
\centerline{\includegraphics[width=\textwidth,clip=]{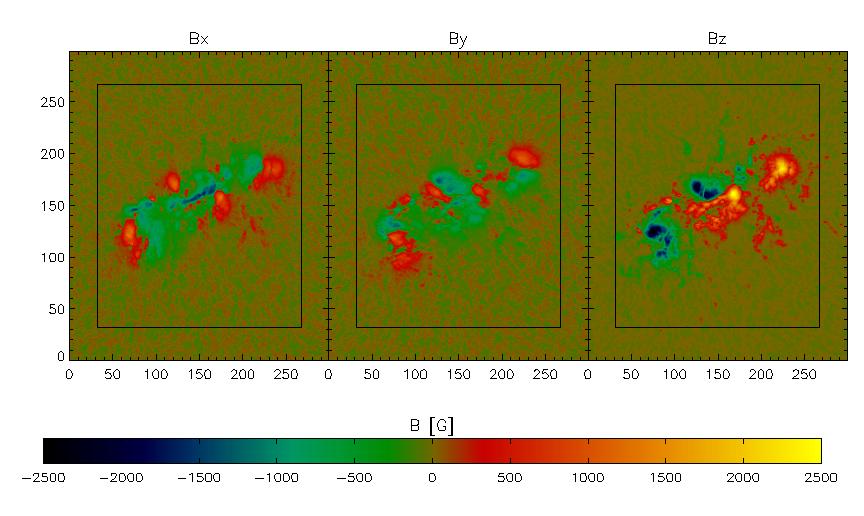}}
 \caption{SDO/HMI vector magnetogram
observed on Feb. 14, 2011 at 20:34UT.
The rectangular area marks the inner box, where $w_f=w_d=1$, see section
\ref{sec3} for details.}
 \label{figure3}
 \end{figure}

The force-free field equations are given by
\begin{eqnarray}
(\nabla \times {\bf B }) \times{\bf B}  & = &  {\bf 0}  \label{forcebal}\\
 \nabla\cdot{\bf B}     & = &   0      \label{solenoidal}
\end{eqnarray}
subject to the boundary condition
\begin{equation}
\bf{B}   =  {\bf B}_{\rm obs} \; {\rm on \, the \, bottom \, boundary}
 \label{Bobs}
\end{equation}
where ${\bf B}$ is the 3D magnetic field  and ${\bf B}_{\rm obs}$
the measured magnetic field vector in the photosphere.
\cite{bineau72,amari:etal06} investigated the mathematical
structure of these equations regarding existence, uniqueness and
well-posedness.
\cite{boulmezaoud:etal00} proofed the existence of force-free
solutions for simple and multiple connected domains. \cite{aly05}
proofed uniqueness of force-free fields for a special cylindrical
configuration.
Several methods have been developed to solve these
equations numerically. For reviews see
\cite{sakurai89,aly89,amari:etal97,wiegelmann08} and within the last
few years the different numerical codes have been intensively
tested, evaluated and compared in
\cite{schrijver:etal06,metcalf:etal08,schrijver:etal08}. As result
of a joint study  \cite{derosa:etal09} it has been concluded that a
successful application of nonlinear force free field (NLFFF)
extrapolation methods require:
\begin{enumerate}
\item Large model volumes at high resolution, which accommodate
most of the magnetic connectivity within an active region and to its surroundings.

The field of view of the isolated active region AR11158, as
shown in Figs. \ref{figure1} and \ref{figure2}, looks like a
suitable candidate to fulfill this requirement.
\item Accommodation of measurement uncertainties in the transverse field
component. 

This has been implemented in recent updates of different NLFFF
extrapolation codes, see
\cite{wheatland:etal09,wiegelmann:etal10a,amari:etal10,wheatland:etal11,tadesse:etal11}.
\item Preprocessing of the photospheric vector field for a
realistic approximation of the upper-chromospheric, nearly
force-free field
\footnote{Preprocessing of inconsistent boundary data is in particular important
for methods using the magnetic field vector directly as boundary condition.
Grad Rubin methods use the normal magnetic field and electric current
(for one polarity) as boundary condition. The vertical current is derived
from the transverse magnetic field and these conditions are per construction
well posed, even if the photospheric magnetic field vector is not force-free.
Consequently preprocessing is not crucial for these methods.}

As we will see in section \ref{quality} the HMI-vector
magnetogram shown in Fig. \ref{figure3} is almost force-free.
Tools for implementing the measurement errors (previous item)
can also deal with the remaining small forces. For comparison we
investigate also preprocessed data. 
\item Force-free models should be compared  with coronal observations.

In Figs. \ref{figure4}, \ref{figure5} and Table \ref{table1} we
compare the force-free models with coronal images observed with
SDO/AIA. 
\end{enumerate}
\section{Instrumentation and data set}
The HMI instrument \cite{schou:etal11} on SDO observes the full Sun
at six wavelengths in the Fe I 6173 \AA $\,$ absorption line.
Filtergrams with a plate scale of $0.5^{\prime\prime}$ pixels are
collected and converted to observable quantities like Dopplergrams,
continuum filtergrams, line-of-sight and vector magnetograms. For
vector data, each set of filtergrams takes 135 seconds to be
completed and the filtergrams are then averaged over a period of
about 12 minutes. [See the HMI-hompage for details:
\url{http://hmi.stanford.edu/}]  To generate the vector
magnetograms, Stokes parameters are derived from the averaged
filtergrams and inverted with the help of a Milne-Eddington
algorithm, an advanced version of the Very Fast Inversion of the Stokes Vector (VFISV)
\cite{borrero:etal10}. The magnetic filling factor is set to be 1 in the inversion.
With the amount of information HMI provides about the line profile, this
provides the most stable results. One could  determine the
filling factor in strong field regions, where it is expected to be close to unity.
In weak field regions, however, it is difficult to resolve the filling factor
as well as field strength.
The $180^{\circ}$ azimuthal ambiguity in the
transverse field is resolved by an improved version of the minimum
energy algorithm \cite{metcalf94,metcalf:etal06,leka:etal09}.
HMI vector field uncertainties depend on field strength, disk position,
and orbital velocity. Formal uncertainties due to the inversion are
computed for each pixel as part of the normal processing.
Conservatively, the random errors in the line-of-sight component are
about 5 G, while the uncertainty in the transverse field is as much as
200 G in weak field regions and as little as 70 G where the field is
strong. The zero point uncertainty in the longitudinal direction is
$<0.1 G$. Additional uncertainties arise because of the disambiguation and
systematic errors that are not as well quantified.

Regions of interest (ROIs) containing strong magnetic fluxes are
automatically identified  \cite{turmon:etal10}. Fig. \ref{figure3}
shows a vector magnetogram, containing AR 11158 observed on Feb. 14,
2011 at 20:34UT. 
After correcting for projection effects \cite{gary:etal90} the data
have been mapped to a local Cartesian coordinate using Lambert equal
area projection \cite{calabretta:etal02}. [An overview about
processing of HMI vector magnetograms can be found at
\url{http://jsoc.stanford.edu/jsocwiki/VectorMagneticField/}.] For
the magnetic field extrapolation, we bin the data to 720 km pixel
(about $1^{\prime\prime}$) and use a computational box of $300
\times 300 \times 160$
grid points. 
\subsection{Quality of the HMI vector magnetogram}
\label{quality} To serve as suitable lower boundary condition for a
force-free modeling, vector magnetograms have to be approximately
flux balanced and the net force and net torque have to vanish.
\cite{wiegelmann:etal06} introduced three dimensionless parameters,
the flux balance $\epsilon_{\mbox{\rm flux}}$, net force balance
$\epsilon_{\mbox{\rm force}}$ and net torque balance
$\epsilon_{\mbox{\rm torque}}$:
\begin{eqnarray}
\epsilon_{\mbox{flux}} &=&
\frac{\int_{S} B_z }{\int_{S} | B_z |  } \nonumber \\
\epsilon_{\mbox{\rm force}} &=&
\frac{|\int_{S} B_x B_z | + |\int_{S} B_y B_z |+
|\int_{S} (B_x^2+B_y^2)-B_z^2  |}
{\int_{S} (B_x^2+B_y^2+B_z^2) }  \nonumber \\
\epsilon_{\mbox{\rm torque}} &=&
\frac{|\int_{S} x ((B_x^2 + B_y^2)-B_z^2) |+
|\int_{S} y ((B_x^2 + B_y^2)-B_z^2)|+
|\int_{S} y  B_x B_z-x  B_y B_z|}
{\int_{S} \sqrt{x^2+y^2} \; (B_x^2+B_y^2+B_z^2) } \nonumber
\end{eqnarray}
where the integrals in $\epsilon_{\mbox{\rm force}}$ and
$\epsilon_{\mbox{\rm torque}}$ correspond to the Maxwell stress
tensor and it's first moment, respectively ( see
\cite{molodensky69,molodensky74,aly89}). For perfectly force-free
consistent boundary conditions these three quantities are zero,
while for real observed data this is hardly the case. For practical
computations, however, it is sufficient these quantities become
small, e.g. $(\epsilon_{\mbox{flux}},\epsilon_{\mbox{\rm force}},
\epsilon_{\mbox{\rm torque}} \ll 1)$. In the following table we list
the values for the used HMI-data-set in the first row. All three
quantities are well below unity, which gives us some confidence that
the data might serve as suitable boundary condition for a force-free
modeling.
$ $ \\\\
\begin{tabular}{lrrr}
Data set & $\epsilon_{\mbox{flux}}$&$\epsilon_{\mbox{\rm force}}$&
$\epsilon_{\mbox{\rm torque}}$\\
\hline
HMI, Feb.14 2011   & $0.0034$ & $0.0564$ & $0.0535$ \\
preprocessed HMI & $0.0037$ & $0.0002$ & $0.0009$ \\
\hline
SFT  Oct.26 1992    & $0.0854$ &$0.6842$ &$0.8837$ \\
Hinode Dec.12 2006 & $0.0167$ &$0.2727$ &$0.3387$ \\
SOLIS Jun.07 2007 & $0.0124$ &$0.6400$&$0.6691$\\
\hline
\end{tabular}
$ $\\\\
To deal vector magnetogram data being inconsistent with the
force-free assumption, we developed a preprocessing routine
\cite{wiegelmann:etal06}, which derives suitable boundary conditions
for force-free modeling from the measured photospheric data.
Applying this procedure to HMI reduces $\epsilon_{\mbox{\rm force}}$
and $\epsilon_{\mbox{\rm torque}}$ further significantly (second
row). It is notable that the values of about $0.05$ in
$\epsilon_{\mbox{\rm force}}$ and $\epsilon_{\mbox{\rm torque}}$ for
the original HMI-vector magnetogram (first row) is significantly
lower as observed for vector magnetograms from other ground based
and space-born missions (rows 3-5) like the Solar Flare Telescope
(SFT), Hinode and SOLIS \footnote{The values refer to other Active
Regions and dates and the values are meant as some typical
value-range for a particular instrument. It is planned to compare
vector magnetograms for one particular active region and time
observed with different instruments (SOLIS and HMI) and the
corresponding force-free models (Thalmann et al., in preparation).
Further investigations are necessary to clarify whether the good
fulfillment of the force-free consistency criteria here is a
property of this particular active region or if the HMI-measurements
are more force-free generally.}. For detailed investigations of
these data-sets see
\cite{wiegelmann:etal06,schrijver:etal08,thalmann:etal08a},
respectively.
\section{Nonlinear force-free field modeling}
\label{sec3}
We solve the force-free equations (\ref{forcebal}) - (\ref{Bobs}) by
an optimization principle as proposed by \cite{wheatland:etal00} and
extended by \cite{wiegelmann04,wiegelmann:etal10a} in the form:
\begin{eqnarray}
L = \int_{V} w_f \frac{|( \nabla \times {\bf B}) \times {\bf B}|^2}{B^2} + w_d \,|\nabla \cdot
{\bf B}|^2 \, d^3V
\nonumber\\
+ \nu \int_{S} ({\bf B} - {\bf B}_{obs})\cdot{\bf W}\cdot({\bf B} - {\bf B}_{obs}) d^2S
\label{defL}
\end{eqnarray}
where $\nu$ is a Lagrangian multiplier which controls the injection speed
of the boundary conditions. $w_f$ and $w_d$
are weighting functions, which are 1
 in the region of interest (inner $236 \times 236 \times 128$ physical box)
 and drop to zero in a $32$ pixel boundary layer towards the lateral and top
 boundaries of the full $300 \times 300 \times 160$ computational domain.
${\bf W}$ is a space-dependent diagonal matrix the element of which
are inverse proportional to the estimated squared measurement error of
the respective field component.
In principle one could compute ${\bf W}$ from the measurement noise and errors
obtained from the inversion of measured Stokes profiles to field components.
Until these quantities become available, a
reasonable assumption is that the magnetic field is measured in strong
field regions more accurately than in the weak field and that the error
in the photospheric transverse field is at least one order of magnitude
higher as the line-of-sight component. Appropriate choices to optimize
 $\nu$  and ${\bf W}$ for use with SDO/HMI magnetograms are investigated
 in this paper.
For a detailed description of the current code implementation and tests
we refer to \cite{wiegelmann04} for the basic code and
\cite{wiegelmann:etal10a} for a description and tests of slow boundary
 injection  and the consideration of measurement errors. For the first
 time we combine the above described algorithm with a multiscale
 approach as described in \cite{wiegelmann08}. For this work we apply our
 code with a 3 level multiscale approach to a SDO/HMI data set with
 $300 \times 300$ points in x and y and extrapolate 160 pixel in hight z.
\subsection{Quality of the reconstructed 3D fields}
To evaluate how well the force-free and divergence-free condition
are satisfied by the reconstructed 3D fields, we monitor a number of
expressions, which are:
\begin{eqnarray}
L_1 &=& \int_{V}  \frac{|( \nabla \times {\bf B}) \times {\bf B}|^2}{B^2}
 \, d^3V \label{defL1} \\
L_2 &=& \int_{V} |\nabla \cdot {\bf B}|^2 \, d^3V \label{defL2} \\
\sigma_j &=& \left( \sum_i \frac{|J_i \times B_i |}{B_i} \right) /
\sum_i J_i \label{defsigmaj} \\
L_{1 \infty} &=&
\sup_{{\bf x} \in V} |{\bf j } \times {\bf B} | \\
L_{2 \infty} &=&
\sup_{{\bf x} \in V} |\nabla \cdot {\bf B}|
\end{eqnarray}
where $L_1$ and $L_2$ correspond to the first and second term in
equation (\ref{defL}), respectively, with the difference that the
integral is carried out in the inner $236 \times 236 \times 128$
physical box, where $w_f \equiv w_d \equiv 1$, excluding the buffer
boundary of $32$ pixel towards the lateral and top boundary of the
computational box.
For potential and linear force-free fields the values correspond to
the discretisation error. Also investigated in the inner box is the
sine of the current weighted average angle $\sigma_j$  between the
magnetic field and electric current (see
\cite{wheatland:etal00,schrijver:etal06} for details) and
 $L_{1 \infty}$ and $L_{2 \infty}$, which are the $L_{\infty}$ norms for
the Lorentz force and divergence, respectively.
\subsection{Code setup and choice of free parameters}
Before we perform nonlinear force-free extrapolations we
use the vertical component
$B_z$ of the HMI-data to compute a potential and a linear force-free field
($\alpha L=2.5, \; \alpha=1.16 \cdot 10^{-8} m^{-1}$)
with a Fourier transform method \cite{alissandrakis81}
Here, $\alpha$ is the linear force-free parameter ($\nabla \times
{\bf B }= \alpha {\bf B} $), which is 0 for a potential field. For a
linear force-free field we calculate, as suggested by
\cite{hagino:etal04}, an averaged value $\alpha=\sum {\mu_0 J_z {\rm
sign}(B_z)}/ \sum{|B_z|}$, where $
J_z=\frac{1}{\mu_0}\left(\frac{\partial B_y}{\partial
x}-\frac{\partial B_x}{\partial y}\right)$ is the vertical current
in the photosphere. %

\begin{figure}
\mbox{\includegraphics[width=6cm,clip=]{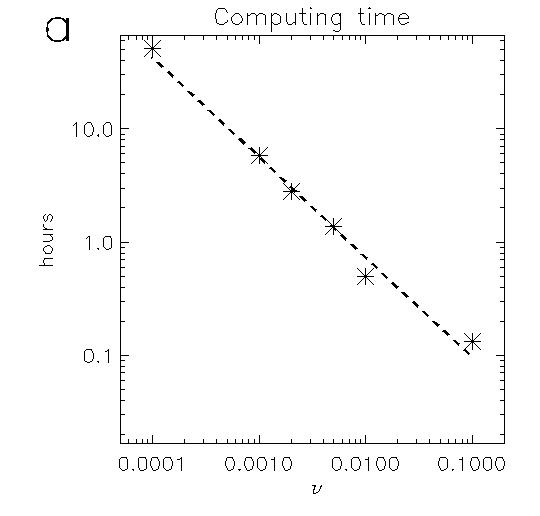}
\includegraphics[width=6cm,clip=]{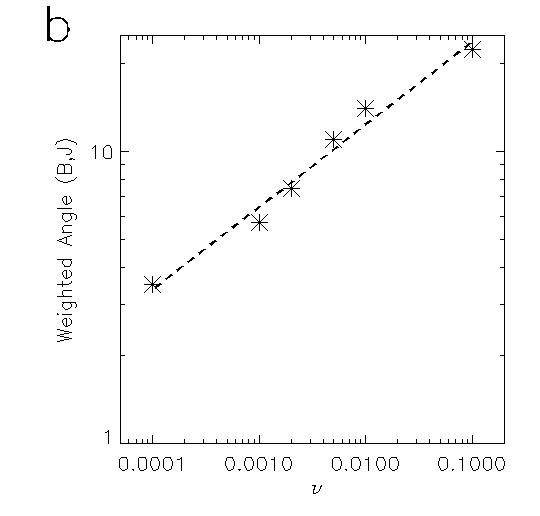}}
\mbox{\includegraphics[width=6cm,clip=]{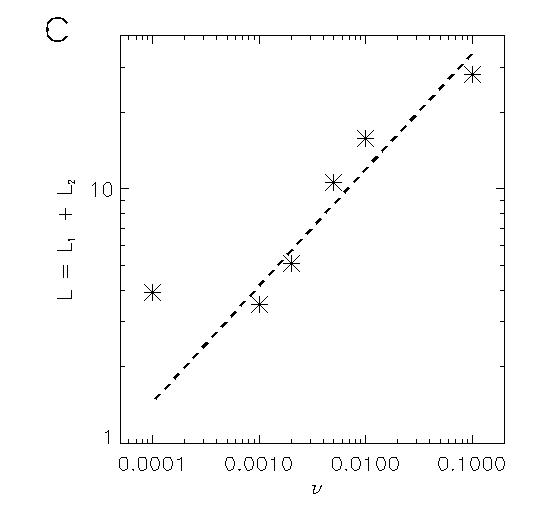}
\includegraphics[width=6cm,clip=]{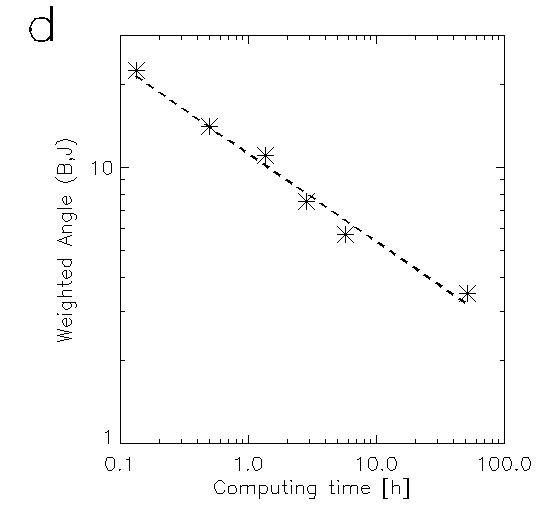}}
\caption{Relation between the Lagrangian multiplier $\nu$ and
computing time (panel a), $\sin^{-1} \sigma_j$ (panel b), $L=L_1+L_2$ (panel c).
Panel d shows the relation of computing time in hours to $\sin^{-1} \sigma_j$.
The values correspond to NLFFF-cases A-F in Table \ref{table1}. The dashed
lines show power law fits in all panels (in panel c the lowest value
$\nu=0.0001$ has been excluded for the power law fit.)}
\label{tablefig}
\end{figure}

For nonlinear force-free fields we minimize the functional
\ref{defL}, we vary the Langrangian multiplier $\nu$ and the Mask
$W$, which we want to optimize. %
For cases A-H we choose $W=B_T/{\rm max}(B_T)$. This seems to be a
reasonable choice as the measurement error in the transverse field
is higher in weak field regions. We vary the Langrangian multiplier
$\nu$ between $0.1$ and $0.0001$ for cases A-F. The lower the value
of $\nu$, the slower the observed boundary becomes injected, so that
the code has more time to relax towards a force-free state. The
computing time 
increases with a power law when $\nu$ decreases $({\rm
Time [h]} \sim 0.013 h \cdot \nu^{-0.88})$, see Fig. \ref{tablefig}a
and column ten in table \ref{table1}). The relation between $\nu$
and $\rm{asin}(\sigma_j)$,  can be be approximated also by a power
law $(\rm{asin}(\sigma_j)\, [\rm degree] \sim 45.7^{\circ} \cdot
\nu^{0.28})$, see Fig. \ref{tablefig}b). The values for force- and
divergence-freeness $L_1$ and $L_2$ are slightly higher for
$\nu=10^{-4}$ than for $10^{-3}$, but the general trend
is that $L_1$ and $L_2$ decrease with decreasing $\nu$ in form of a
power law, $L \sim 98 \cdot \nu^{0.46}$, with $L=L_1+L_2$, see Fig.
\ref{tablefig}c). It seems that the choice $\nu=0.001$ is optimal,
as higher values of $\nu$ correspond to worse fulfilment of all
force-free consistency criteria $(\sigma_j, L_1, L_2)$ and a lower
$\nu$ only increases the computing time drastically, but does not or
hardly improve the solution.

In cases G and H we investigate the influence of preprocessing on
the result. We used a standard-preprocessing parameter set
$\mu_1=\mu_2=1, \mu_3=0.001, \mu_4=0.01$. These parameters control
the amount of force-freeness, torque-freeness, nearness to the
actually observed  data and smoothing, respectively (see
\cite{wiegelmann:etal06} for details on preprocessing).
NLFFF-extrapolations have been carried out here for $\nu=0.01$ and
$\nu=0.001$ and we find that $L_1$ and $L_2$ are smaller for
extrapolations from preprocessed data. Similar as in the unprocessed
case $L_1, L_2, \sigma_j$ decrease with decreasing $\nu$, while the
computing time increases. The computing time for preprocessed data
is about a factor of a three ($\nu=0.01$) or two ($\nu=0.001)$
higher compared with the unprocessed cases. The angle between
magnetic field and current $\rm{asin}(\sigma_j)$ does not improve,
however, and becomes worse (factor $1.3$) for $\nu=0.001$. A reason
for the L-values becoming lower, while $\sigma_j$ remains the same,
might be that some strong current-peaks are smoothed out by
preprocessing. If we consider that without preprocessing (case E)
 $L_1$ and $L_2$ are already of the
order of the discretisation error of the potential field, the lower value
$\sigma_j$ and the shorter computing times, we conclude that preprocessing is
not necessary for this data set.

In the cases I-L we investigate the effect of different
mask-functions. We choose a unique mask in cases I and J, which
means that we do not consider different errors in high and low field
strength regions in the photospheric vector magnetogram. As one can
see (in comparison with cases B and E) all three force-free
consistency criteria are worse and consequently one should not use a
unique mask. Interesting are the cases K and L, where we choose the
mask $W=\left( B_T/{\rm max}(B_T) \right)^2$. This choice gives more
weight to strong than to weak regions, similar as in the linear
cases A-H, but prefers strong regions significantly more. For
$\nu=0.01$ (case K compared with B) $L_1$ and $L_2$ are better and
$\sigma_j$ worse. For $\nu=0.001$ (case L compared with E) all three
criteria are fulfilled somewhat better for the quadratic mask
function. The computing time is, however, a factor of $1.6$ longer
for the quadratic case (L).
We conclude, that if $\nu$ is
sufficiently low, than the final equilibrium is quite robust
regarding the exact choice of the mask function profile.

For comparison we also extrapolated the magnetogram with the old code
version, which uses a fixed lower boundary and does not contain a mask
and Lagrangian multiplier (cases M and N without
and with preprocessing, respectively).
Without preprocessing, all three force-free consistency criteria are
fulfilled worse as for the new code, because the old code has no possibility
to correct for inconsistencies in the magnetograms. Preprocessing improves
the result for $L_1$ and $L_2$ by about a factor of four, but $\sigma_j$
hardly improves. The computing time for the old code (case N) is, however,
significantly lower (factor of 5, 9 compared with E,L, respectively)
as for the best runs with the new code. If we compare results of the old code
(case N, after preprocessing) with results from the new code with similar
computing times (cases C and G) the performance are similar.
Higher computing times are the price we have to pay to get better force-free
consistent equilibria.

In column nine in Table \ref{table1} we present the ratio of the
total magnetic energy to the energy of a potential field $E/E_0$.
While the correct value is a priori unknown, this criteria cannot
serve directly as a quality measure of the reconstructed
NLFFF-fields, except that $E/E_0$ should be greater than unity.
$E/E_0$ is an important quantity, as it defines an upper limit for
the free magnetic energy, which could become converted in kinetic
and thermal energy during eruptions. Taking the average of all 14
NLFFF-models we find $E/E_0=1.20 \pm 0.06$, and if we consider only
the best cases, say where $\rm{asin}(\sigma_j)\, < \, 10^{\circ}$
(cases D,E,F,H,J,L) one finds $E/E_0=1.24 \pm 0.03$. For long time
series it will probably not be possible to extrapolate all
magnetograms with several different parameter sets, but we propose
to do this for a some magnetograms within a time series to derive an
error-estimation for $E/E_0$.  We find that this quantity is not
significantly influenced by the chosen parameter set (value of
Lagrangian multiplier $\nu$, mask function profile $W$,
preprocessing) if the force-free consistency criteria $L_1,L_2,
\sigma_j$ are fulfilled and one should check them for each
extrapolation from vector magnetograms. We find that the
$L_{\infty}$ norms for Lorentz force and divergence behave very
similar as the integral forms. Therefore both norms can alternatively
be used to evaluate the quality of the extrapolated field.
%
%
\begin{sidewaystable}
\caption{Evaluation of force-free field models. The first column names the model
and in column 2 and 3 are shown the used model parameters. Column 4-8 show
different force-free consistency evaluations, column 9 the magnetic energy compared
with a potential field and column 10 the computing time on one processor.
Finally, in the last three column we compare magnetic field lines in different regions
(denoted red, white, green also in Fig. \ref{figure5}) with an AIA-image.}
$ $ \\
\begin{tabular}{llrrrrrrlrlll}
Data set & $\nu$&Mask&$L_1$&$L_2$&$\sin^{-1} \sigma_j $ &
$L_{1 \infty}$&$L_{2 \infty}$
& $E/E_0$
\footnote{Preprocessing smoothes also somewhat the vertical magnetic field component
and the energy of a corresponding potential field is $3\%$ lower as for
computations from the unprocessed $B_z$. We provide also the ratio to
the original potential field in brackets for these cases.}& Time
&AIA-red & AIA-white & AIA-green\\
\hline
pot. field &&&$2.0$&$1.9$&$52.5^{\circ}$&1198&239&$1.00$&
&$ 473 \pm 66 $&$ 265 \pm 9 $&$ 754 \pm 102 $\\
LFF $\alpha2.5$ &&&$1.9$&$1.8$&$19.3^{\circ}$&1312&220&$1.04$&
&$ 898 \pm 270 $&$ 241 \pm 5 $&$ 858 \pm 163 $\\
\\
A) NLFFF &$0.1$&$\propto |B_T|$& $17.4$&$10.8$&$22.4^{\circ}$&3682&472&$1.14$& 8min
&$ 165 \pm 13 $&$ 207 \pm 4 $&$ 645 \pm 75 $\\
B) NLFFF &$0.01$&$\propto |B_T|$& $10.7$&$4.9$&$14.0^{\circ}$&3545&356&$1.17$&30min
&$ 185 \pm 11 $&$ 211 \pm 9 $&$ 451 \pm 86 $\\
C) NLFFF &$0.005$&$\propto |B_T|$& $7.6$&$3.0$&$11.0^{\circ}$&3013&312&$1.20$&1h:22min
&$ 200 \pm 9 $&$ 216 \pm 18 $&$ 411 \pm 40 $\\
D) NLFFF &$0.002$&$\propto |B_T|$& $3.8$&$1.4$&$7.5^{\circ}$&2004&247&$1.23$&2h:49min
&$ 213 \pm 11 $&$ 232 \pm 20 $&$ 398 \pm 28 $\\
E) NLFFF &$0.001$&$\propto |B_T|$& $2.5$&$1.0$&$5.7^{\circ}$&1769&210&$1.25$&5h:46min
 &$233 \pm 16$&$221 \pm 5 $& $425 \pm 35$\\
F) NLFFF &$0.0001$&$\propto |B_T|$& $2.8$&$1.1$&$3.5^{\circ}$&1771&209&$1.26$&51h:32min
  &$ 244 \pm 23 $&$ 209 \pm 35 $&$ 432 \pm 67 $\\
G) Prepro &$0.01$&$\propto |B_T|$& $3.5$&$1.5$&$14.0^{\circ}$&1020&126&$1.25 (1.22)$&1h:33min
&$ 207 \pm 22 $&$ 234 \pm 12 $&$ 429 \pm 41 $\\
H) Prepro &$0.001$&$\propto |B_T|$& $1.2$&$0.4$&$7.6^{\circ}$&736&70&$1.28 (1.24)$&10h:06min
&$ 234 \pm 29 $&$ 285 \pm 18 $&$ 384 \pm 27 $\\
\\
I) NLFFF &$0.01$&$ 1$& $19.6$&$8.3$&$17.7^{\circ}$&3682&438&$1.10$&24min
&$ 167 \pm 9 $&$ 204 \pm 12 $&$ 550 \pm 86 $\\
J) NLFFF &$0.001$&$ 1$& $13.0$&$5.7$&$8.7^{\circ}$&2593&330&$1.18$&3h:30min
&$ 201 \pm 18 $&$ 215 \pm 24 $&$ 313 \pm 54 $\\
K) NLFFF &$0.01$&$\propto B_T^2$& $7.0$&$4.7$&$14.6^{\circ}$&2317&324&$1.08$&9min
&$ 171 \pm 8 $&$ 201 \pm 7 $&$ 499 \pm 95 $\\
L) NLFFF &$0.001$&$\propto B_T^2$& $1.7$&$0.8$&$4.7^{\circ}$&1656&187&$1.24$&9h:13min
&$ 267 \pm 13 $&$ 206 \pm 17 $&$ 449 \pm 53 $\\
\\
\multicolumn{3}{l}{M) Old code}& $27.3$&$11.5$&$16.9^{\circ}$&4591&581&$1.15$&46min
&$ 169 \pm 13 $&$ 209 \pm 15 $&$ 598 \pm 75 $\\
\multicolumn{3}{l}{N) Old code, preprocessed}&$6.5$&$2.9$&$16.6^{\circ}$&1196&185&$1.20 (1.17) $&1h:02min
&$ 194 \pm 25 $&$ 239 \pm 7 $&$ 525 \pm 98 $\\
\end{tabular} \\
\label{table1}
\end{sidewaystable}

\section{Comparison with AIA-images}
\begin{figure}
\mbox{\includegraphics[width=5cm,clip=]{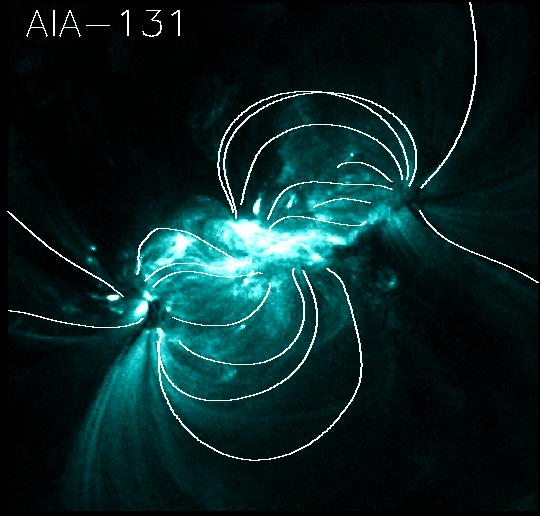}
\includegraphics[width=5cm,clip=]{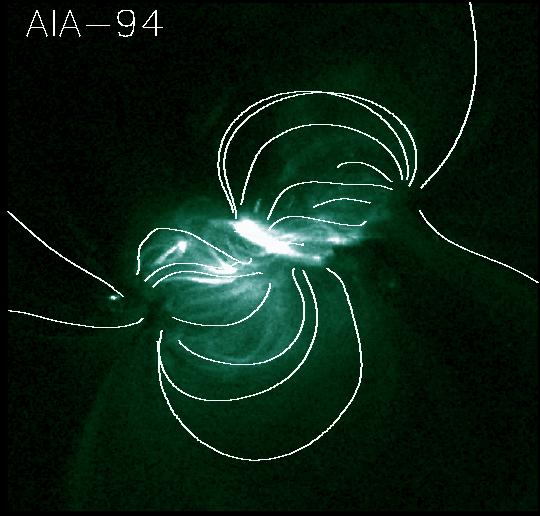}
\includegraphics[width=5cm,clip=]{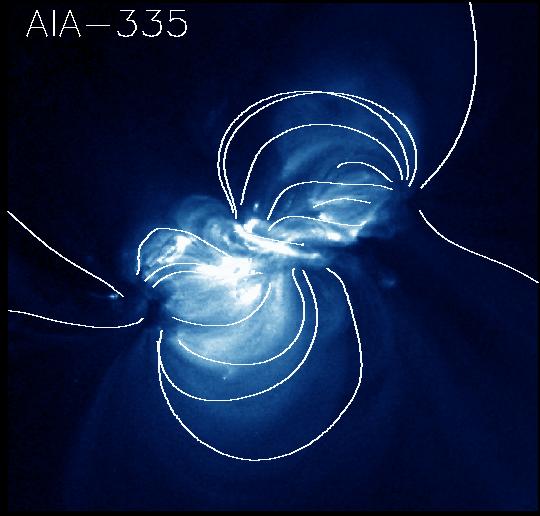}}
\mbox{\includegraphics[width=5cm,clip=]{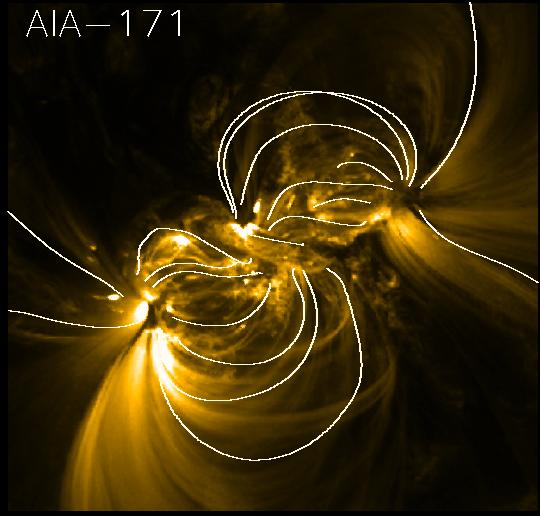}
\includegraphics[width=5cm,clip=]{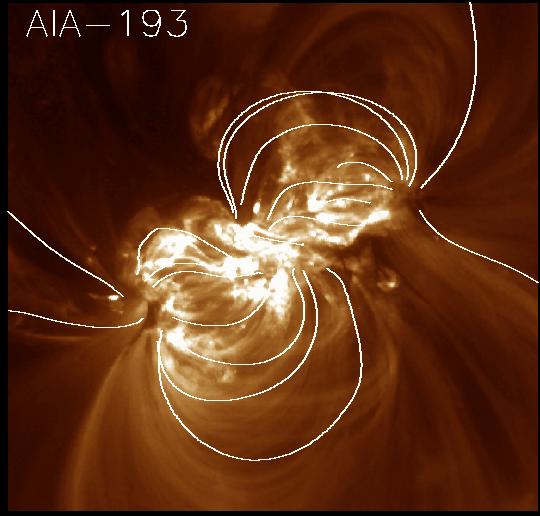}
\includegraphics[width=5cm,clip=]{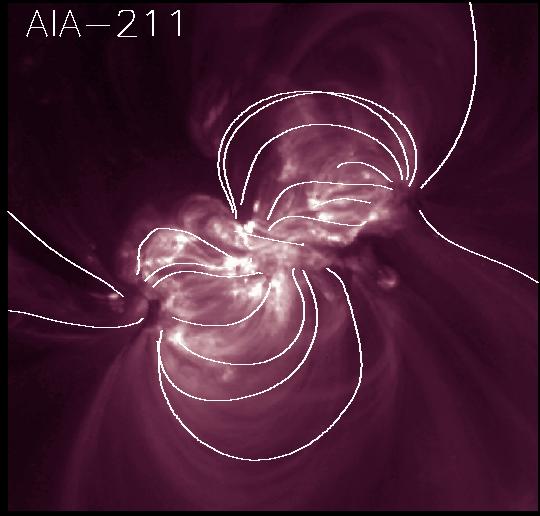}}
\mbox{\includegraphics[width=5cm,clip=]{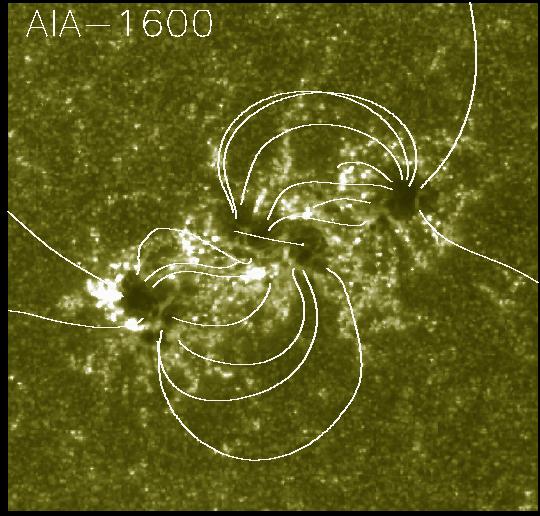}
\includegraphics[width=5cm,clip=]{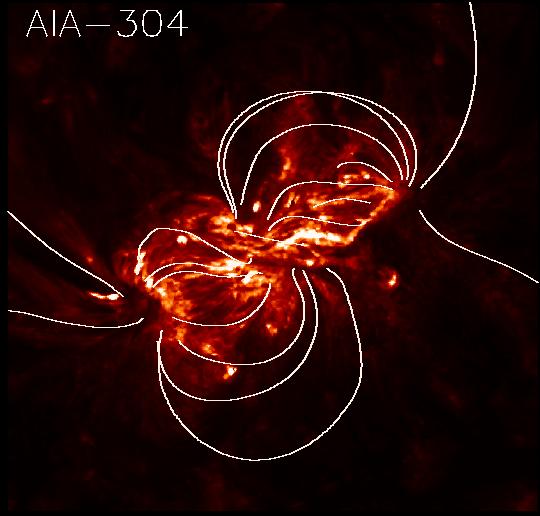}
\includegraphics[width=5cm,clip=]{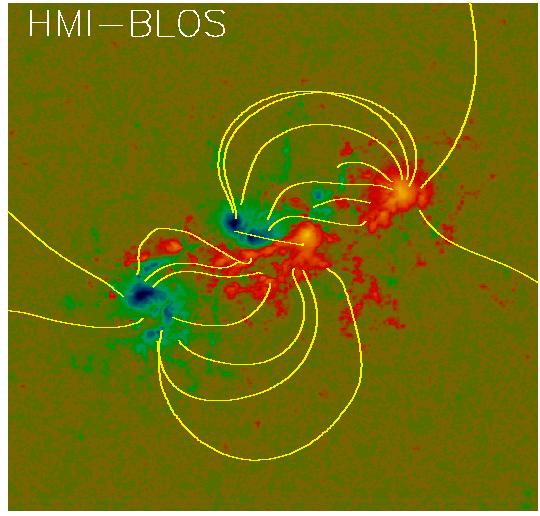}}
\caption{AIA-images of AR11158 in different wavelengths,
observed on Feb. 14, 2011 at 20:34UT.
Over-plotted are some selected field lines from NLFFF
model E.}
\label{figure4}
\end{figure}
\begin{figure}
\mbox{\includegraphics[width=6cm,clip=]{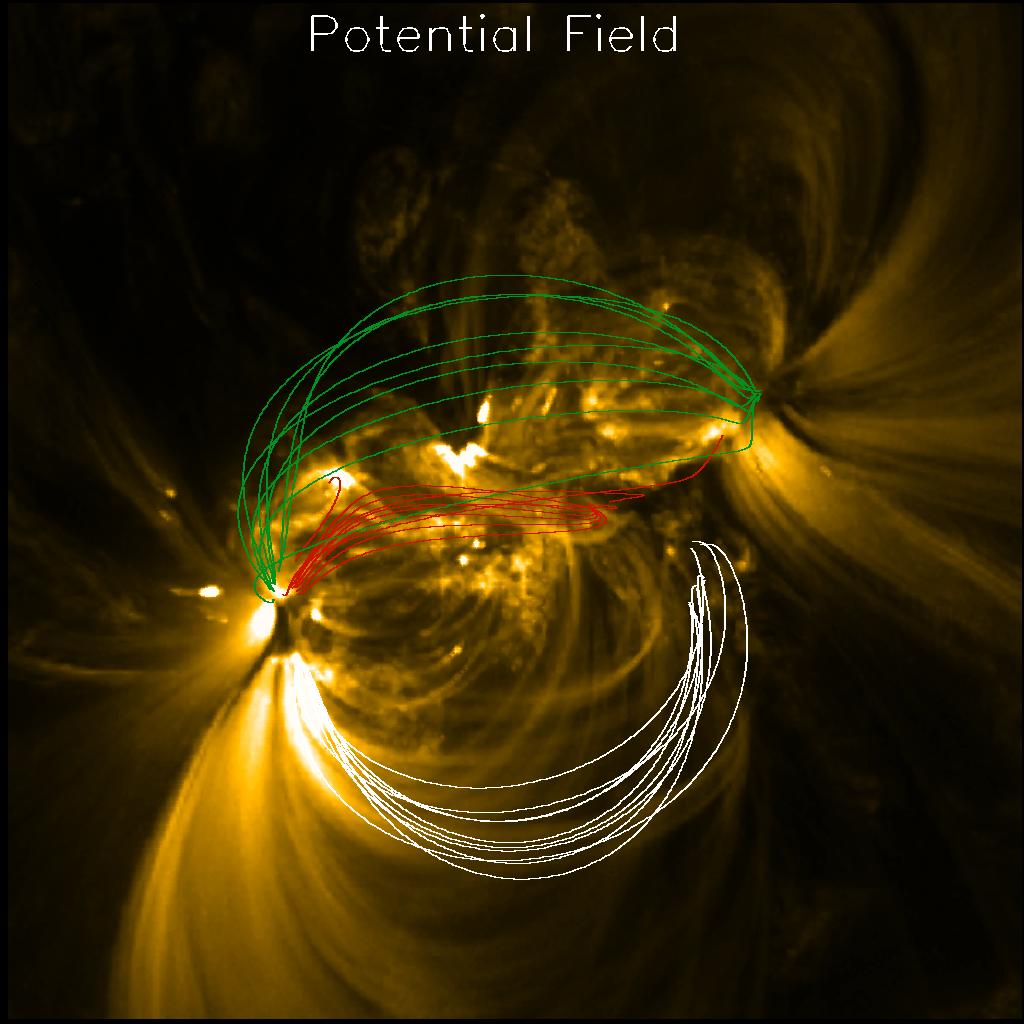}
\includegraphics[width=6cm,clip=]{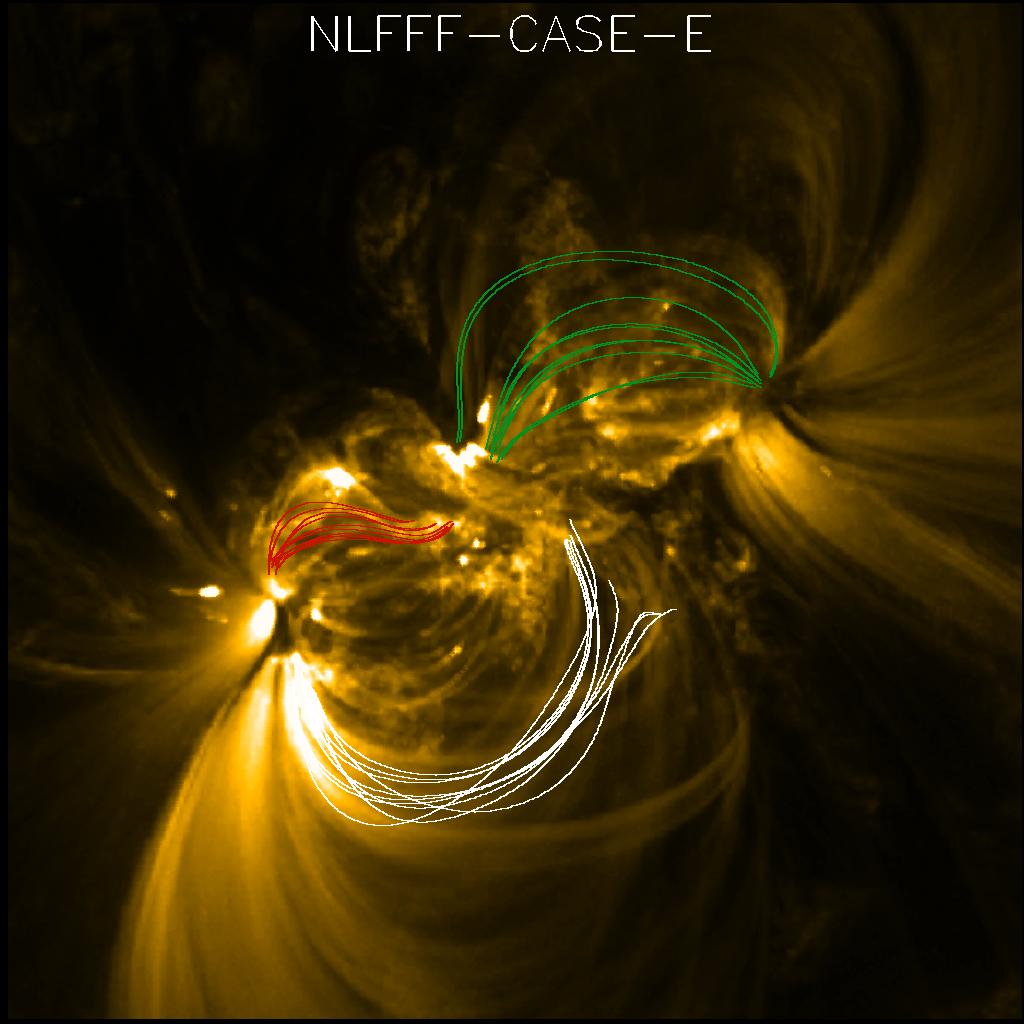}}
\mbox{\includegraphics[width=6cm,clip=]{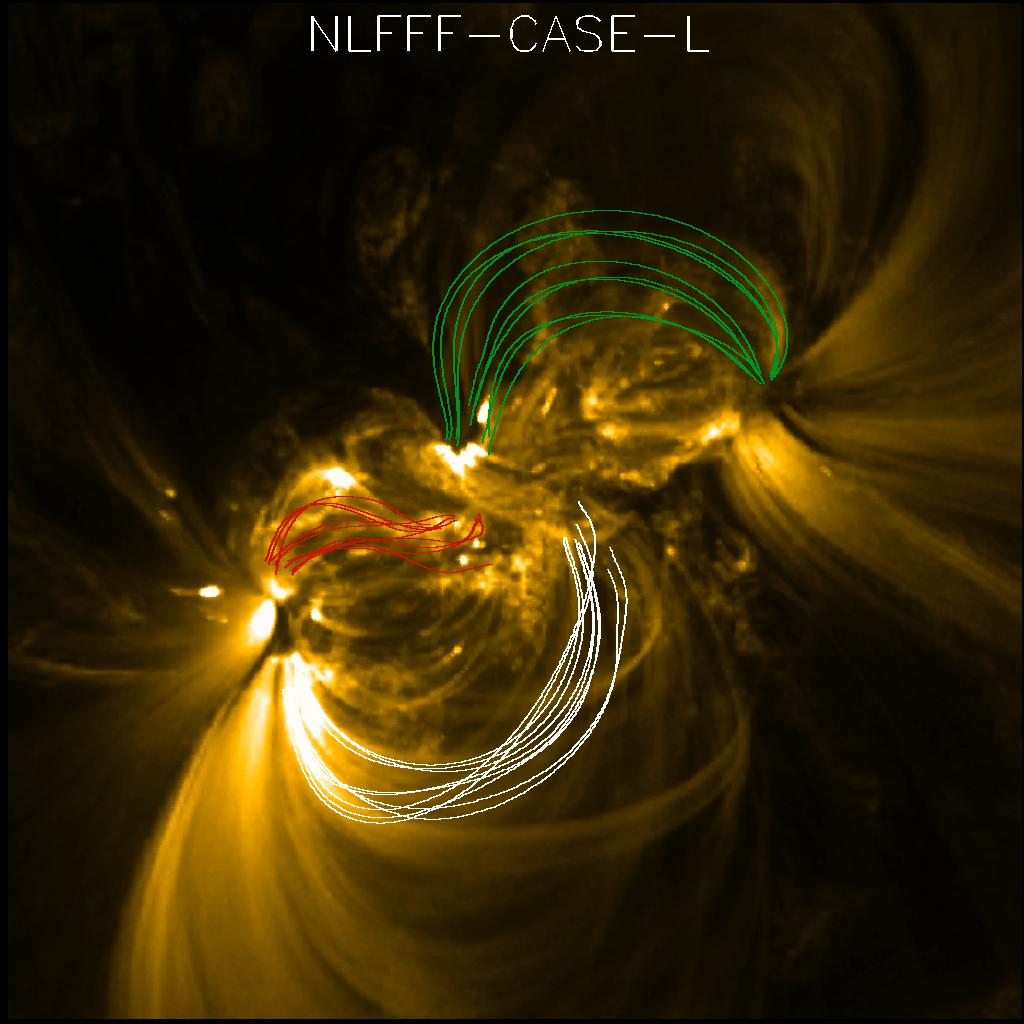}
\includegraphics[width=6cm,clip=]{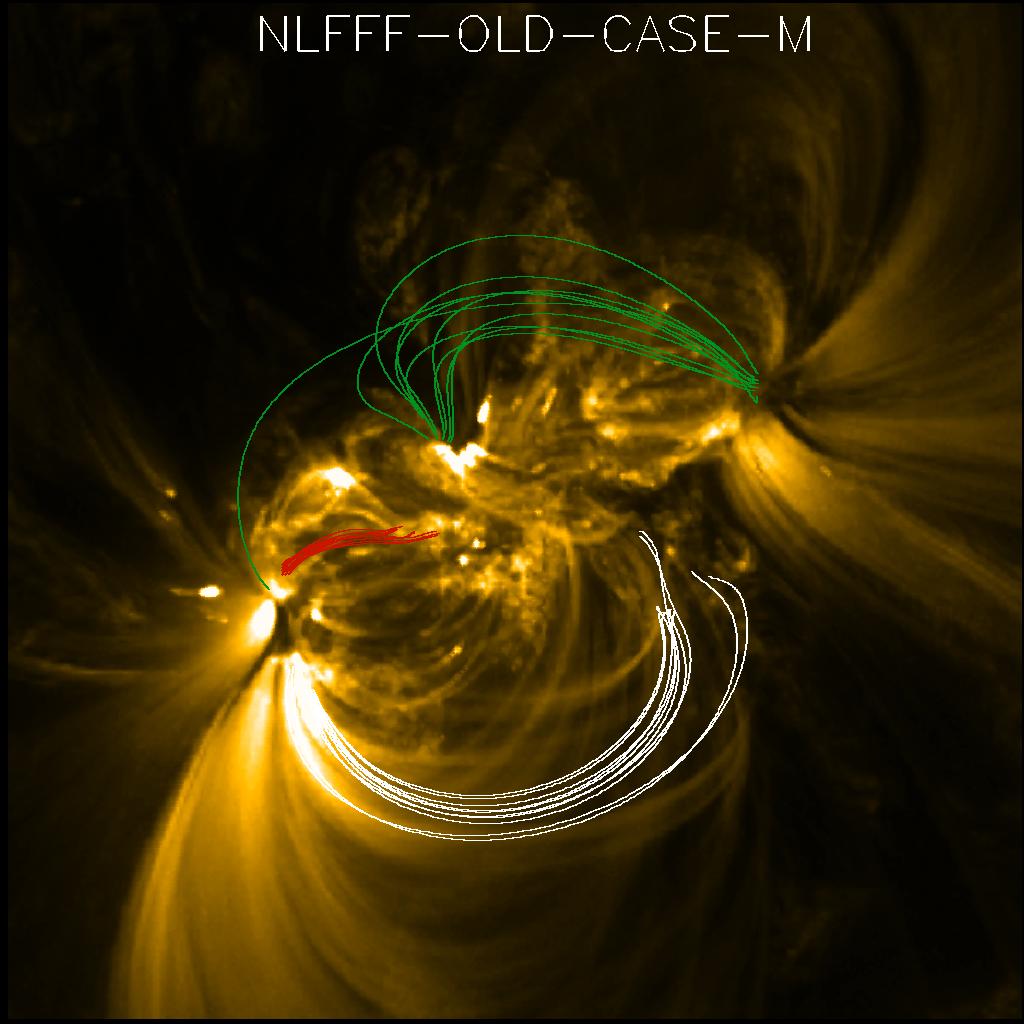}}
\caption{AIA-171 \AA \, images with field lines as computed
from selected models. The red, white and green loops here
correspond to AIA-red, AIA-white and
AIA-green in table \ref{table1}}
\label{figure5}
\end{figure}
As pointed out in \cite{derosa:etal09} force-free field models
should be compared with coronal observations in order to quantify to
which extend they correctly reproduce the coronal magnetic field
configuration. In Fig. \ref{figure4} we show some arbitrary chosen
force-free field lines (from NLFFF-model-E) in comparison with
AIA-images in different wavelengths. See \cite{lemen:etal11} for an
overview on AIA. Qualitatively it looks like the field lines
reasonably agree with the observed plasma-loops. Deviations,
however, are also clearly recognizable. In the following we aim to
estimate the difference between force-free field lines and plasma
loops quantitatively. While stereoscopic reconstructed loops in 3D
as used in \cite{derosa:etal09,conlon:etal10} are not available, we
compare our results with an AIA-image from one viewpoint. A basic
assumption is that the plasma is frozen into the magnetic field,
hence the plasma loops outline the magnetic field lines.
Consequently, the gradient of the intensity  parallel to the
magnetic field lines should be small. We are  interested to detect
bright loops (high intensity) and to quantify the deviation of
projected field lines (computed from the NLFFF-models) and plasma
loops (as visible in AIA-images) as
\begin{equation}
C=\frac{\sqrt{\int_S (\nabla I(s))^2} ds}{\int_S I(s) ds},
\end{equation}
where $I(s)$ is the intensity along a projected magnetic field line
\footnote{For convenience
the values in last three columns in table \ref{table1}
have been multiplied by $10^5$ and rounded
to get 3-figure integer numbers.}.
This criterion finds loops of high $I(s)$ with small intensity
gradients along the loops (low $\nabla I(s)$). For other
possibilities to define the penalty function $C$ see
\cite{wiegelmann:etal05,conlon:etal10}. 

For a quantitative comparison, we compute a number of field lines,
originating $\pm 5$ pixels around previously chosen locations. These
locations are $P_1=(80,30), P_2=(85,100)$ and $P_3=(225,185)$. Field
lines not closing within the extrapolation domain are excluded from
the quantitative comparison and the ten field lines owing the lowest
values of $C$ are considered for further analysis. The average and
standard deviation of $C$  for these ten field lines in each region
are displayed in the last three columns of table \ref{table1},
respectively. The field lines origination from $\pm 5$ pixels around
$P_1, P_2$ and $P_3$ are shown as red, white and green curves,
respectively for some of the models in Fig. \ref{figure5}.
The field lines corresponding to the potential field model, the
NLFFF-models E and L (which best performed
 for the force-free consistency criteria, see last section) and a the old
 (fixed boundary, model M) NLFFF-code are shown in  Fig.
 \ref{figure5}.

The white loops in Fig. \ref{figure5} seem to agree reasonably well
with AIA for all models, even those of the potential field model.
The average penalty function for these loops is in the range 200-300
for all models. The red S-shaped coronal loops on the other hand
cannot be identified with a potential or linear-force free model the
(the linear model performing even worse than the potential one). All
NLFFF-models show a much better agreement, with the penalty function
for NLFFF-model (about 165-265) being in most cases less than half
as large as for the potential field. The penalty function for the
green loops is higher for all models, but the penalty function for
the best NLFFF-models is about a factor of two better than for the
potential field. It seems that these green loops are the most
challenging loops to be reconstructed and the old NLFFF-code, which
used a fixed boundary for the transverse potential field, performs
only slightly better as the potential field. The best NLFFF-models
(in the sense of most force-free, in particular cases E and L shown
in Figure \ref{figure5}), however, indicate the correct field
topology also for the green loops. Using the penalty function $C$ to
evaluate the quality of the reconstruction clearly favors
NLFFF-models over potential and linear FF models. This criterion is
not sensible enough, however, to definitely favor one of the NLFFF-models. In
the future one should consider to sophisticate the comparison of
magnetic field models with coronal images, e.g., by applying
different penalty functions, use  loops-structures extracted from
the images and doing comparisons in different AIA-wavelengths. For
some earlier observation from SDO (for which vector magnetograms
have not been released yet) one could also compare the results with
images taken from vantage points with one or both of the
STEREO-spacecraft or compare them directly with stereoscopic
reconstructed 3D-loops. This cannot, however, be implemented as a
standard diagnostic for NLFFF-models, as the angle between the two
STEREO spacecraft and SDO becomes to large for stereoscopy.

\section{Conclusions and Outlook}
Within this work we carried out nonlinear force-free coronal field
extrapolations of an isolated active region  based on data from
SDO/HMI. The vector magnetogram is almost perfectly flux balanced
and the field of view was large enough to cover also the weak field
surrounding the active region. Both conditions are necessary  in
order to carry out meaningful force-free computations. We also
found, that the photospheric magnetogram satisfied well the
force-free criteria. The net force and torque are considerably
smaller than in 
earlier measurements of other active region fields with SFT, Hinode
and SOLIS. We do not know for sure, however, if this is a general
property of HMI, or only true for this particular isolated active
region. A comparison of
an active region measurement with different instruments is planned.
The data could be used directly as boundary conditions for nonlinear
force-free field computation and preprocessing the photospheric
field was not necessary. The new code version incorporates errors of
the measurement, in particular in the transverse field, and injects
the boundary data slowly, controlled by a Lagrangian multiplier
$\nu$. The error incorporation is controlled by using a
mask-function, which is $1$ for the most thrustworthy data and $0$
where one cannot thrust the data. Unless an exact error computation
becomes available from inversion and ambiguity removal of the
photospheric magnetic field vector, a reasonable assumption is that
the field is measured more accurately in strong field regions and we
carried out computations with the mask $\propto B_T$ and $\propto
B_T^2$. For a sufficient small Lagrangian multiplier $\nu=0.001$ we
found that the resulting coronal fields are force and divergence
free in the sense that the remaining residual forces are of the
order of the discretisation error of potential and linear force-free
fields. The weighted angle between magnetic field and electric
current is about $5^{\circ}-6^{\circ}$. The resulting field is
almost identical for both masks, but computations with the $\propto
B_T^2$ mask take significantly longer (9h:13min instead of 5h:46min
for the $\propto B_T$ mask. Injecting the boundary faster by
choosing a higher Lagrangian multiplier (say $\nu=0.01$) speeds up
the computation (to half an hour), but the residual forces are
higher and current and field are not well aligned. Inserting the
boundary even slower (say $\nu=0.0001$) leads to much longer
computing times (more then 50h), but does not improve the solution.
We conclude that the choice $\nu=0.001$ and a mask $\propto B_T$ or
$\propto B_T^2$ are the optimal choices for this data set. The
computations have been carried out on one processor on a Linux-PC.
Our code has been parallelized with Open-MP, but rather then
processing a single magnetogram with a parallelized code, it is
planned to process different magnetograms of a time series
simultaneously. While the time cadence of HMI vector magnetograms is
about 12min, NLFFF computations for one magnetogram on one processor
take about 6h. Consequently the requirement is about 50 processors
(for each active region) in order for the NLFFF-tools to catch up
with the data stream from HMI.

An important question is to which extend the optimum parameters for
this data set can also be applied to other active regions from
SDO/HMI. A key point is to monitor the consistency criteria of the
magnetogram as well as the remaining residual forces, alignement of
fields and currents in the reconstructed 3D field. A comparison of
the magnetic field model with AIA-images should also always be done.
We used the $171 \AA$ channel here, because loops are well visible in
this wavelength. The question, how coronal magnetic field models
can be validated best by coronal observations should be, however,
further be investigated. Magnetic field lines are 3D-structures
and many field lines might not be filled with plasma and thus
not visible in EUV-images.
Also it is not trivial how/if one can use the different wavelength in AIA
to validate coronal field models.

A stumbling stone for AR-NLFFF models could be that other active regions
are not so well isolated
as AR11158 investigated here, but might be magnetically connected to other
ARs and the quiet Sun. Such situations require
full disk vector magnetograms and force-free computations in spherical
geometry, as for example carried out from full-disk SOLIS-measurements in
\cite{tadesse:etal11a}. Due to it's very nature extrapolations from full
disk magnetograms have to be calculated with a reduced spatial resolution or
one has to accept significant longer computing times. Global force-free
coronal magnetic field models can also be used to specify the lateral
boundaries for active region modelling for non-isolated ARs.
%
%

%

%

%

%

%
\begin{acks}
 $ $ \\
Data are courtesy of NASA/SDO and the AIA and HMI science teams.
We are gratefully to Marc DeRosa for his help with AIA-data.
This work was supported by  by DLR grant 50 OC 0904 and
DFG grant WI 3211/2-1.
\end{acks}

%
%
\bibliographystyle{spr-mp-sola-cnd} 
\bibliography{tw}
%
%
%
%

\end{article}
\end{document}